\documentclass{ptephy_v1}%%%%%% to generate preprint number with ptep logo

\preprintnumber{XXXX-XXXX} %%% %%% Insert preprint number here

\usepackage{multirow}

\begin{document}

\title{Development of Convolutional Neural Networks for an Electron-Tracking Compton Camera}

%%%% To generate auto affiliation numbers please use \author{}\affil{} command

\author[1]{Tomonori Ikeda}
\author[1]{Atsushi Takada}
\author[1]{Mitsuru Abe}
\author[1]{Kei Yoshikawa}

\author[1]{Masaya Tsuda}
\author[1]{Shingo Ogio}
\author[1]{Shinya Sonoda}
\author[2]{Yoshitaka Mizumura}
\author[1]{Yura Yoshida}

\author[1]{Toru Tanimori}
%\author[1]{Hidetoshi Kubo}
%\author[3]{Shunsuke Kurosawa}
%\author[4]{Kentaro Miuchi}
%\author[5]{Tatsuya Sawano}
%\author[6]{Kenji Hamaguchi}

\affil[1]{Graduate School of Science, Kyoto University, Kitashirakawa Oiwakecho, Sakyo, Kyoto, Kyoto, 606-8502, Japan \email{ikeda.tomonori.2s@kyoto-u.ac.jp}}
\affil[2]{Scientific balloon, Institute of Space and Astronautical Science, Japan Aerospace Exploration Agency, Yoshinodai 3-1-1, Chuou, Sagamihara, Kanagawa, 252-5210, Japan}
%\affil[3]{Institute of Materials Research, Tohoku University, Katahira 2-1-1, Aoba, Sendai, Miyagi, 980-8577, Japan}
%\affil[4]{Graduate School of Science, Kobe University, Rokkoudai, Nada, Kobe, Hyogo, 657-8501, Japan}
%\affil[5]{Graduate School of Natural Science and Technology, Kanazawa University, Kakuma, Kanazawa, Ishikawa, 920-1192, Japan}
%\affil[6]{Department of Physics, University of Maryland, Baltimore County, 1000 Hilltop Circle, Baltimore, MD 21250, USA}

%%% To include the collaborator name... Please use the command "\collaborator"
%%% For example: \collaborator{ATLAS Collaboration}

%%%%%%%%%%%%%%%%%%%%%%%%%%%%%%%%%%%%%%%%%%%%%%%%%%%%%%%%%%%%%%%%%
% ABSTRACT
%%%%%%%%%%%%%%%%%%%%%%%%%%%%%%%%%%%%%%%%%%%%%%%%%%%%%%%%%%%%%%%%%
\begin{abstract}%
Electron-tracking Compton camera, which is a complete Compton camera with tracking Compton scattering electron by a gas micro time projection chamber, is expected to open up MeV gamma-ray astronomy.
The technical challenge for achieving several degrees of the point spread function is the precise determination of the electron-recoil direction and the scattering position from track images. 
We attempted to reconstruct these parameters using convolutional neural networks.
Two network models were designed to predict the recoil direction and the scattering position.
These models marked 41~degrees of the angular resolution and 2.1~mm of the position resolution for 75~keV electron simulation data in Argon-based gas at 2~atm pressure.
In addition, the point spread function of ETCC was improved to 15~degrees from 22~degrees for experimental data of 662~keV gamma-ray source.
These performances greatly surpassed that using the traditional analysis.
%Improved ETCC is expected to expand to reach of the MeV gamma-ray sky.

\end{abstract}

\subjectindex{MeV gamma-ray, MPGD, TPC, Deep learning}

\maketitle

%%%%%%%%%%%%%%%%%%%%%%%%%%%%%%%%%%%%%%%%%%%%%
% INTRO
%%%%%%%%%%%%%%%%%%%%%%%%%%%%%%%%%%%%%%%%%%%%%
\section{Introduction}
Gamma-ray observations in the low-energy gamma-ray band from a hundred keV to several MeV address various astrophysical phenomena; the nucleosynthesis and the explosion mechanism in the supernovae~\cite{Matz,Diehl2003,Churazov2015}, the particle acceleration process in the active galactic nuclei or the gamma-ray bursts~\cite{Chiaberge2001,Ballo2002,Briggs1999,Preece2000} and electron-positron annihilation line in the Galactic center region~\cite{OSSE,Siegert,COSI}.
However, the observation in this energy band remains poorly explored compared to that of X-ray, GeV and TeV band, known as ``MeV gap". 
Even now, COMPTEL~\cite{COMPTEL}, onboard {\it Compton Gamma Ray Observatory} launched in 1991, is the most sensitive observation of the MeV sky.
The causes of stagnant in MeV observation are huge gamma-ray backgrounds from the atmosphere (albedo) and generated in the instruments by cosmic-ray interactions. 
In addition, conventional Compton cameras like COMPTEL have unclean images according to the Compton circle due to the lack of the direction information of the recoil electron. Such pseudo imaging is not capable of background rejection and does not keep quantitative imaging~\cite{TANIMORI2017}.
Thus, it was pointed out that conventional Compton cameras need the additional parameters of the recoil direction of the Compton electron in order to reduce backgrounds~\cite{SCHONFELDER2004193}.

Electron-Tracking Compton Camera (ETCC) is a complete Compton camera with tracking recoil electrons. It can record all information on Compton kinematics for overcoming the problem of conventional Compton cameras. 
Hence ETCC determines a unique incident gamma-ray direction~\cite{TANIMORI2017}.
%the point-spread function (PSF) based on the optical principle against the conventional Compton camera.
%In addition, it achieves great background rejection power. 
A key device of ETCC is the tracking detector to detect the recoil electron since the point-spread function (PSF) is highly dependent on the determination accuracy of the recoil direction and the scattering position~\cite{Tanimori2015}. 
Thus, the SMILE group (Sub-MeV gamma-ray Imaging Loaded-on-ballon Experiment) developed the gas time projection chamber (TPC) based on $\mu$-PIC~\cite{Takada}, which is one of the micro-pattern gas detectors. Although the gas TPC gives us the precise track information for low energy electrons, these electrons are influenced by the multiple scattering and make complex track images.
%The development of the analysis method in order to determine scattering position and direction with a high precision is needed.
The technical challenge is the development of the analysis method with an exact determination of the scattering position and the recoil direction from these track images.

The feature extraction from images is the field of computer vision. 
Neural Networks (CNNs) have achieved great success in image classifications.
%Mostly, Convolutional Neural Networks (CNNs) were a great success in image classification.
Since 2010, many architectures like AlexNet~\cite{AlexNet}, VGG~\cite{VGG}, GoogLeNet~\cite{GoogLeNet}, and ResNet~\cite{ResNet} were proposed in the ImageNet Large Scale Visual Recognition Challenge~\cite{ImageNet} and achieved dramatic progress.
In particle physics and astrophysics experiments, modern machine learning techniques have been actively applied and developed.
There are several purposes of signal-background recognition, particle identification, and event reconstruction. Machine learning showed significant performance improvements than the traditional way based on domain knowledge. In the gas TPC application, the CNN was developed for the polarization extraction from photoelectron track images taken with X-ray polarimeters and improved the polarization sensitivity by 10\%--20\%~\cite{KITAGUCHI}. 
The NEXT experiment, searching for neutrinoless double-beta decay, utilized CNN to identify electron-positron pair production from the topological signature. Then the signal efficiency was improved compared to non-CNN-based analysis~\cite{NEXT}.
In our case, CNN would also be a promising approach.

%In this paper, we improve the performance of ETCC by employing CNNs.
In this paper, we describe the design of CNNs to predict the scattering position and the electron-recoil direction from track images taken by the TPC. 
In addition, we evaluate the imaging performance of the simulation and experiment data, and compare with a traditional way.

%In Section~\ref{sec:ETCC}, we introduce the ETCC of SMILE-2+ and show recoil-electron image obtained by tracking detector. In addition, key variables of the Compton camera are defined here.
%In Section~\ref{sec:CNN}, we describe designs of two CNNs to predict the scattering position and the recoil direction. These perfomance are showed using unseen simulation data in Section~\ref{sec:}.
%Finally, we evaluate the PSF 
%The goal of this paper is to be 

%%%%%%%%%%%%%%%%%%%%%%%%%%%%%%%%%%%%%%%%%%%%%
% ETCC
%%%%%%%%%%%%%%%%%%%%%%%%%%%%%%%%%%%%%%%%%%%%%
\section{Electron-Tracking Compton Camera}\label{sec:ETCC}
\begin{figure}[!h]
\centering
\includegraphics[width=6in]{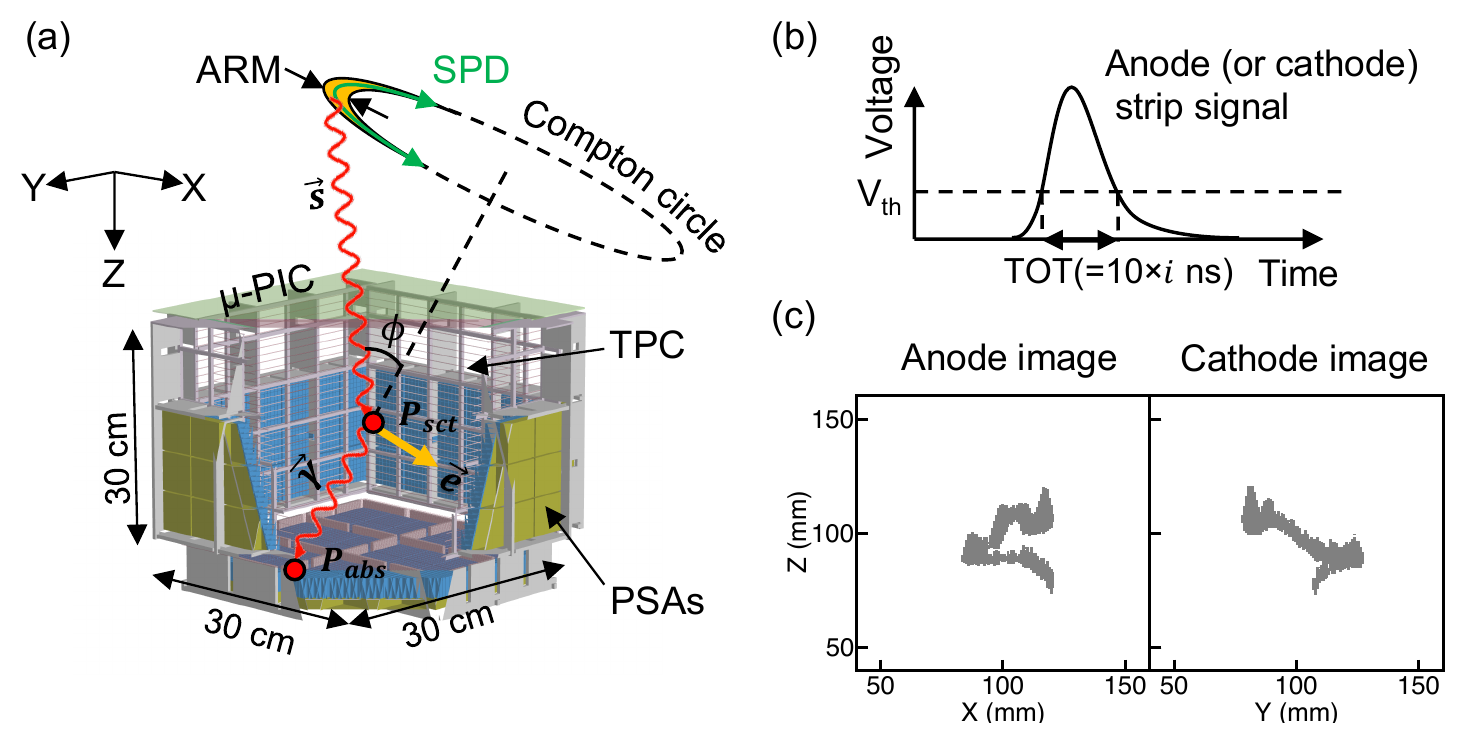}
\caption{(a) Schematic view of the ETCC on the SMILE-2+ experiment. The incident gamma-ray coming from the source location $\vec{r}_{\gamma}$ undergoes Compton scattering into the TPC filled with argon-based gas. The energy, scattering position $P_{\rm sct}$ and direction $\vec{r}_{e}$ of the recoil electron are reconstructed by the TPC based on $\mu$-PIC. 
Finally, the scattering gamma-ray $\vec{r}_{\gamma '}$ is absorbed at $P_{\rm abs}$ with PSAs.
(b) Schematic image of the TOT.
(c) Track images of recoil electron with 160~keV from $\mu$-PIC anode and cathode strips. %Gray pixels indicate TOTs with 10 ns clocks.
}
\label{fig:ETCC}
\end{figure}

The dominant interaction process of MeV gamma-rays and materials is the Compton scattering.
The schematic view of the Compton kinematics is described in Fig.~\ref{fig:ETCC}.
The conventional Compton camera obtains the information of scattering gamma-ray energy $E_{\gamma}$, the absorption position of scattering gamma-ray $P_{\rm abs}$, the scattering position of gamma-ray $P_{\rm sct}$, and the electron-recoil energy $K_{\rm e}$ other than the electron-recoil direction $\vec{e}$. Then, the scattering angle $\phi$ is written as follows,
\begin{equation}
\cos\phi=1-\frac{m_{\rm e}c^{2}}{E_{\gamma}+K_{e}} \frac{K_{\rm e}}{E_{\gamma}}.
\end{equation}
The conventional Compton cameras are insufficient to resolve the kinematic equation of Compton scattering due to the lack of the electron-recoil direction.
While the incident gamma direction $\vec{s}$ is only reconstructed as a Compton circle, the overlapping of many Compton circles determines the source location. 
In the conventional Compton camera, the PSF is only defined by the angular resolution measure (ARM). The ARM means the angular distance from the reconstructed Compton circle to the known source location or the determination accuracy of Compton scattering angle $\phi$. The ARM depends on the determination accuracies of the absorption point and the scattering point of gamma-ray, and the energy resolution.

COMPEL was incredibly successful via Compton imaging techniques. However, the achieved sensitivity was modest. 
%The most important things we learned from COMPTEL are that huge gamma-ray backgrounds are generated in the satellite itself by cosmic rays.
One of the most important things we learned from COMPTEL is that the huge gamma-ray backgrounds are generated in the satellite itself by cosmic rays.
Under such backgrounds, the sensitivity is overestimated if we only use the ARM. 
Sch\"onfelder reported the conceivable background in COMPTEL and concluded that the most sensitive tool to increase the sensitivity is to reduce the instrumental background rate~\cite{SCHONFELDER2004193}. 
He also proposed the measurement of recoil electrons in order to suppress the background. From this effort, we developed ETCC as the next generation Compton telescope.

The ETCC provides all parameters of Compton scattering including the electron-recoil direction by the tracking detector. % and recover the SPD.
Figure~\ref{fig:ETCC}(a) shows a schematic view of the ETCC on the SMILE-2+ experiment, which is our second balloon experiment to observe the celestial objects at the  high altitudes~\cite{Tanimori2020}.
The ETCC completely resolves the Compton kinematics and determines the unique incident gamma-ray direction $\vec{s}$ using the following equation,
\begin{equation}\label{eq:reconstruct}
\vec{s} = \frac{E_{\gamma}}{E_{\gamma}+K_{\rm e}} \vec{\gamma} + \frac{\sqrt{K_{\rm e}(K_{\rm e}+2m_{\rm e} c^{2})}}{E_{\gamma}+K_{\rm }} \vec{e}.
\end{equation}
The ETCC gives the complete PSF from the ARM and the Scattering Plane Deviation (SPD)~\cite{TANIMORI2017}, which is the accuracy of the determining the scattering plane. The ETCC can achieve high background-rejection power by localizing the arrival direction on the Compton circle thanks to the SPD.

%%%%%%%%%%%%%%%%%%%%%%%%%%%%%%%%%%%%%%%%%%%%%%%%%%%%%%%%%%%%%%%%%%
% Detector
%%%%%%%%%%%%%%%%%%%%%%%%%%%%%%%%%%%%%%%%%%%%%%%%%%%%%%%%%%%%%%%%%%
The ETCC of the SMILE-2+ experiment consists of pixel scintillator arrays (PSAs) and a gas TPC based on a micro pattern gas detector $\mu$-PIC~\cite{Takada}.
The 36 PSAs with a thickness of 26~mm and the 18~$\times$~4 PSAs with a thickness of 13~mm are arranged bottom and side of TPC, respectively.
Each PSA is made of GSO (Gd$_{2}$SiO$_{5}$:Ce) scintillators of 8~$\times$~8 pixels with a pixel size of 6~$\times$~6~mm$^{2}$. 
PSAs are deployed as absorbers to detect the energy $E_{\gamma}$ and the position  $P_{\rm abs}$ of the scattering gamma-ray.
The TPC is filled with argon-based gas (95\%~Ar~+~3\%~CF$_{4}$~+~2\%~C$_{4}$H$_{10}$ at a pressure of 2~atm) as a Compton-scattering target and has a drift length of 30~cm.
The electron track and energy information was detected by $\mu$-PIC, which has 768~$\times$~768 strips with a pitch of 400~$\mu$m; the detection volume 30~$\times$~30~$\times$~30~cm$^{3}$.
In order to reduce the power consumption, every two strips of $\mu$-PIC are grouped.
Thus, the readout strip pitch is 800~$\mu$m in the SMILE-2+ experiment.

The anode and cathode strips of $\mu$-PIC are linearly arranged in the X and Y directions, respectively. The Z direction is reconstructed by the measured drift velocity and drift time.
Thus two-dimension images (XZ and YZ) are obtained. The strip signals are processed in FE2009bal CMOS ASIC chips~\cite{MIZUMOTO201540} including preamplifiers, shapers, and comparators.
Each amplified signal is compared to the threshold voltage and synchronized  with 100~MHz (10~ns) clocks. 
Finally, the addresses of $\mu$-PIC strips with the time-over-threshold (TOT) are recorded.
Figure~\ref{fig:ETCC}(b) shows the schematic image of the TOT.
%The TOT is the time between rising and falling edges crossing the threshold voltage, which is correlated with the charge information on each strip.
The TOT is the time between rising and falling edges crossing the threshold voltage, which is correlated with the convolution of the charge information and the track-length along the z-axis on each strip.
Since the TOT is measured by the comparator with 100~MHz, it is digitized with 10~ns. 
Figure~\ref{fig:ETCC}(c) shows a typical example of the electron track obtained by $\mu$-PIC. %The gray pixel is corresponding to TOTs. 
The length of gray pixels along the Z-axis on each strip is corresponding to the TOT multiplied by the drift velocity.
%The z coordinate is reconstructed considering the drift velocity of the ionized electron in TPC. 
%TOTs include not only z coordinate information but also charge deposit information.
The low density scatter medium and the fine interval read-out detector $\mu$-PIC provides fine track images. The scattering position $P_{\rm sct}$ and electron-recoil direction $\vec{e}$ are reconstructed from these two images.

The traditional method to determine the scattering position is to utilize the skewness of TOTs, which relies on the fact that the stopping power of electron recoil depends on their residual energy, and the end-point of the track has large TOTs. The skewness is written by the following equation,
\[
S = \frac{\mu_{3}}{\mu_{2}^{3/2}}=\frac{\langle(x-\langle x\rangle)^{3}\rangle}{\langle(x-\langle x\rangle)^{2}\rangle^{3/2}}.
\]
It is represented as the dimensionless ratio between the third and second moments of the TOTs. 
The skewness provides the information on whether the endpoint or the start point is on the right or left side of the image.  
Consequently, the maximum or minimum of track image is adopted as a scattering position. After calculating the scattering position, the recoil direction is determined by the linear fitting algorithm using TOTs within 4~mm of the reconstructed scattering point to prevent multiple scattering effects. This method is suitable for relatively straight tracks, while it is less accurate for curved tracks by the multiple scattering.

%%%%%%%%%%%%%%%%%%%%%%%%%%%%%%%%%%%%%%%%%%%%%
% CNN
%%%%%%%%%%%%%%%%%%%%%%%%%%%%%%%%%%%%%%%%%%%%%
\section{Convolutional Neural Network}\label{sec:CNN}
CNN is one of the artificial neural networks. It has an input layer, hidden layers, and an output layer. The hidden layer is generally composed of a combination of the convolutional layer, the pooling layer, and a fully connected layer. The convolutional layer, which is the most important element in the CNN, automatically extract features from the image by a weighting matrix, called a filter or kernel. 
The pooling layer is used to down-sampling the feature maps extracted in the convolutional layer by taking an average or maximum.
In a fully connected layer, all nodes are connected to all activations in the previous layer. A deep hierarchical structure of the convolutional layer and pooling layer has expressive capacity; a shallow layer learns low-level features or local correlations, and a deep layer learns high-level features or global correlations. Thus, general CNNs are made up of stacks of the convolutional layer and pooling layer.

We implemented two independent network models to predict the scattering position and the electron-recoil direction using the Keras/Tensorflow framework~\cite{Tensorflow}.
The common point in both network models is that the input layer takes two-channel images of  256~$\times$~512 pixels corresponding to 204.8~$\times$~194.56~mm$^{2}$ in XZ and YZ dimensions. Thus, the size of the input parameter ($\phi$, $\cos\theta$) is 2~$\times$~256~$\times$~512 ($=$262144).
In this section, we describe the characteristics of these network models.

%%%%%%%%%%%%%%%%%%%%%%%%%%%%%%%%%%%%%%%%%%%%%%%%%%%%%%%%%%%%%%
% Model for direction
%%%%%%%%%%%%%%%%%%%%%%%%%%%%%%%%%%%%%%%%%%%%%%%%%%%%%%%%%%%%%%
\subsection{Network model to predict the electron-recoil direction}
In 2014, the Visual Geometry Group (VGG~\cite{VGG}) model was received high praise in the classification category of the image recognition competition. This architecture simply increases the number of layers by connecting the convolution layer and the fully connected layer.
The feature is that the reduction of the filter size enables to implement a deeper network and learn high-level features.
We have designed a network model based on the VGG to predict the electron-recoil direction.

\begin{figure}[!h]
\centering
\includegraphics[width=6in]{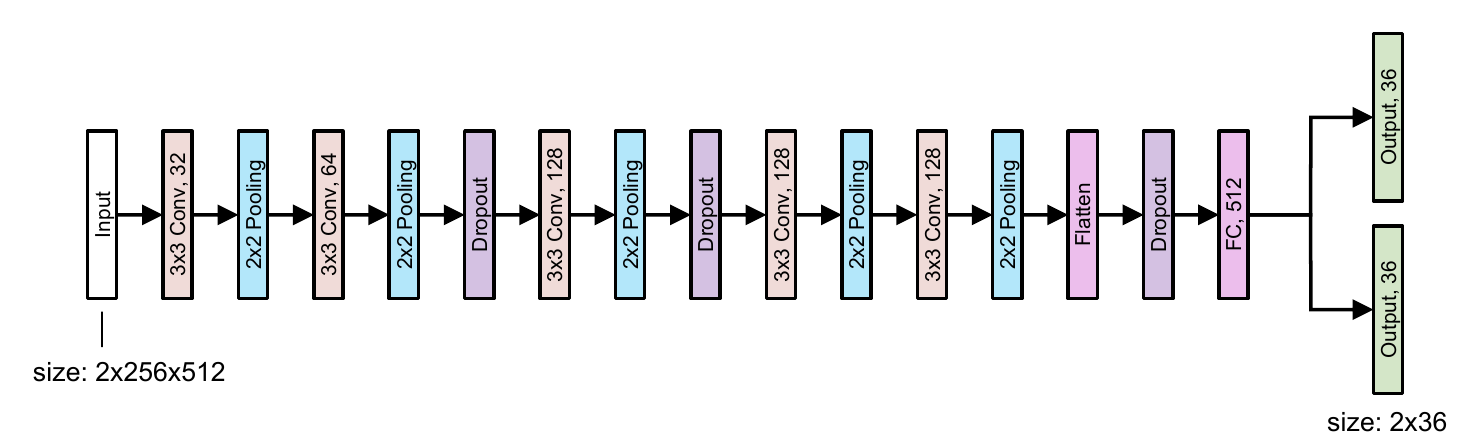}
\caption{Network model to predict the electron-recoil direction. White and green boxes are input and output layer, respectively. Orange boxes represent convolution layers with stride 1, which are denoted as (filter size) Conv, (number of filters).
Blue boxes are maximum pooling layer with the filter size of  2~$\times$~2 and stride 2. 
Purple boxes are dropout layers with the rate set to 20\%.
Pink boxes represent full connected or flatten layers.}
\label{fig:model_dir}
\end{figure}

The schematic view of the network model is shown in Fig.~\ref{fig:model_dir}. 
This model has two output layers to predict the cosine of zenith angle $\cos\theta$ and azimuth angle $\phi$ of electron recoil direction.
The prediction of each angle is divided into 36 classes, each having a width of 0.055 and 10 degrees, respectively.
Thus, the size of the output parameter is 2~$\times$~36.
Output layers are activated by softmax functions to compute each of the probability distribution.
The hidden layer is a hierarchical representation of convolutional layers and pooling layers.
All convolutional layers have a filter size of 3~$\times$~3 and stride 1.
All pooling layers takes the maximum value with the filter size of 2~$\times$~2 and stride 2.
In addition, three dropout layers, in which some number of layer outputs are randomly ignored, are deployed to prevent over-fitting.
In the final stage, two-dimensional feature maps are converted to a one-dimensional vector in flatten layer and connected to the fully connected layer with 512 nodes.
All layers except for pooling layers and output layers are activated by the rectified linear unit (ReLU~\cite{ReLU}). 

%%%%%%%%%%%%%%%%%%%%%%%%%%%%%%%%%%%%%%%%%%%%%%%%%%%%%%%%%%%%%%
% Model for position
%%%%%%%%%%%%%%%%%%%%%%%%%%%%%%%%%%%%%%%%%%%%%%%%%%%%%%%%%%%%%%
\subsection{Network model to predict the scattering position}\label{sec:network_model_dir}
The accuracy of the scattering position is expected to be the pixel size resolution.
While the VGG model is able to learn high-level feature with the deep layer, the detail position information is lost at the same time as the size of feature maps are reduced.
For spatially dense predictions like semantic segmentation, which is the process of separating image into some object of interest with pixel-level, Fully Convolutional Networks (FCN~\cite{FCN}), and U-Net~\cite{U-Net} were developed. 
These architectures have an encoder part followed by a decoder part. 
The encoder part is the typical classifier network like the VGG. 
The decoder part expands the feature map and projects onto high-resolution images.
In order to predict the scattering position, we built the network model upon the U-Net architecture.

\begin{figure}[!h]
\centering
\includegraphics[width=6in]{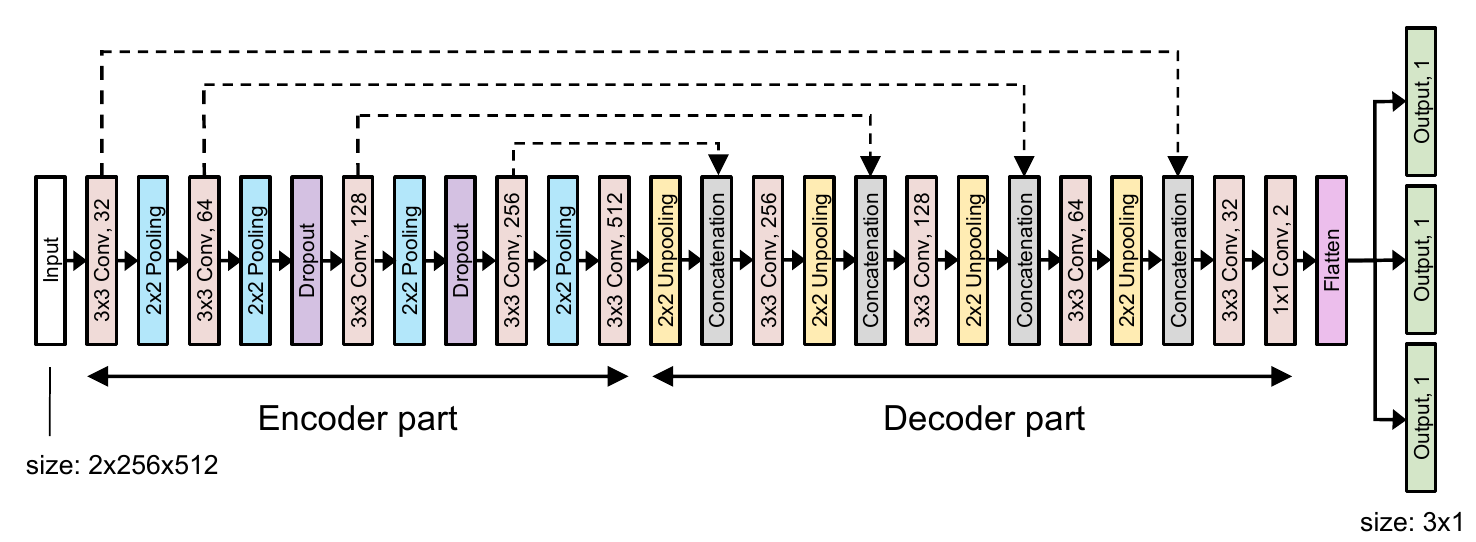}
\caption{Network model to predict the scattering position.
White and green boxes are input and output layer, respectively. 
Orange boxes represent convolution layers with stride 1, which are denoted as (filter size) Conv, (number of filters).
Blue boxes are maximum pooling layers with a filter size of  2~$\times$~2 and stride 2. 
Purple boxes are dropout layers with the rate set to 20\%.
Yellow boxes are unpooling layers.
Pink box describes a flatten layer.
Black dotted arrows indicate concatenation operation to combine the output of the convolution layers in the encoder part to the decoder part.}
\label{fig:model_pos}
\end{figure}

The schematic view of the network model is depicted in Fig.~\ref{fig:model_pos}.
In order to predict the scattering positions ($x$, $y$, $z$), this model has three output layers, which have linear activation functions. The size of the output parameter is 3~$\times$~1. 
The scattering position is predicted on a regression problem.
The hidden layer is comprised of the encoder and decoder part. 
The first encoder section extracts features at different scale sizes with down-sampling as well as the network to predict the electron-recoil direction.
Convolutional layers have 3~$\times$~3 kernesl with stride 1, and pooling layers are maxpooling operation and stride 2.
%In the middle convolutional layer extract feature information with low spatial resolution.
In the second decoder section, pooling layers are replaced with unpooling layers as up-sampling and expands feature maps. 
We deploy the convolution transpose layer as unpooling operation with the filter size of 2~$\times$~2 and stride 2.
In addition, up-sampling images are connected by down-sampling images in concatenation layers and recovered the high spatial resolution. The final up-sampling layer is flattened and connected to output layers.
All layers except for pooling layers and output layers are activated by the ReLU.

%%%%%%%%%%%%%%%%%%%%%%%%%%%%%%%%%%%%%%%%%%%%%
% MONTE CALRO SIMULATION
%%%%%%%%%%%%%%%%%%%%%%%%%%%%%%%%%%%%%%%%%%%%%
\section{Training and validating network via a monte calro simulation}
\subsection{Preparing a training and validating data set}
The training and validating data are prepared by the Monte Carlo (MC) simulation package of the SMILE experiment based on the Geant4~\cite{Geant4}.
In order to reproduce experimental images, the following procedures are taken.
\begin{itemize}
    \item The full detector geometry is constructed in Geant4. Electron beams are generated in TPC volume with the uniform position and direction. Then, the interaction points and the deposit energies are recorded.
    \item From the deposit energies, the number of ionized electrons is calculated considering the fano factor. In addition, the collected positions of ionized electrons on $\mu$-PIC strips are obtained according to the diffusion effect simulated by MAGBOLTZ~\cite{MAGBOLTZ}.
    \item Waveforms of $\mu$-PIC strips are represented by overlapping waveform templates, which are  calculated from the simulated  $\mu$-PIC pulse signal using Garfiled++~\cite{Takada_2013}.
    These waveforms are digitized and  TOTs are obtained by comparing to the threshold. 
    Finally, calculated TOTs are encoded to the experimental data format.
\end{itemize}

We analyze simulated data and make XZ and YZ images of 256~$\times$~512 pixels so that the tracks are in the center of the images like Fig.~\ref{fig:ETCC}(b).
The 630k and 70k MC events for 5--200~keV, which were fully contained in the TPC volume, were generated for the training and validating data sets, respectively. 
Also, each class has almost the same number of samples; one class has about 17500 samples for the training data set.

%%%%%%%%%%%%%%%%%%%%%%%%%%%%%%%%%%%%%%%%%%%%%
% 
%%%%%%%%%%%%%%%%%%%%%%%%%%%%%%%%%%%%%%%%%%%%%
\subsection{Training and validating network}
We trained the networks and optimized learnable parameters by minimizing the loss function between the ground truth and the prediction. Training the model to predict the electron-recoil direction, the ground truth is  expressed by the one-hot vector $\vec{p}$. For example, the $i$-th class constitutes angles $\phi$ ranging from $\pi i/18-\pi$ to $\pi(i+1)/18 - \pi$. Then, the $i$-th label is one ``1" and the rest is zero ``0" in the one-hot vector. We employed the cross-entropy loss, which is a typical loss function in the classification task, represented by $H(p,q)=-\Sigma_{i}p_{i}\log q_{i}$, where $p_{i}$ is the $i$-th label of the ground truth and $q_{i}$ is the predicted one. On the other hand, we employed the mean square root as a loss function to train the model to predict the scattering position. In both cases, an Adam optimizer was used to minimize loss function with a learning rate of 0.0001. A batch size of 4 was chosen. We trained models for ten epochs where validation losses saturated. 
The huge amount of calculations was done by NVIDIA Geforce GTX 2070 super graphics processing units (GPU).

\begin{figure}[!h]
\centering
\includegraphics[width=6in]{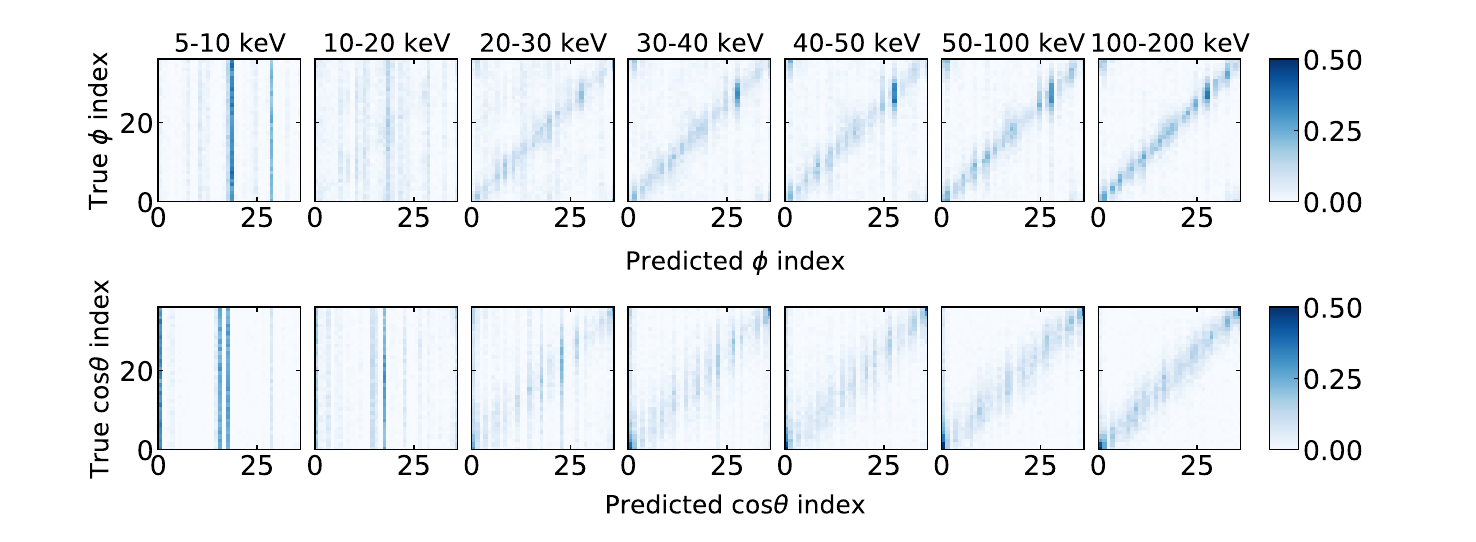}
\caption{Confusion matrices at the $\phi$ (top panel) and $\cos\theta$ (bottom panel) class predictions. }
\label{fig:confusing_matrix}
\end{figure}

\begin{figure}[!h]
\centering
\includegraphics[width=6in]{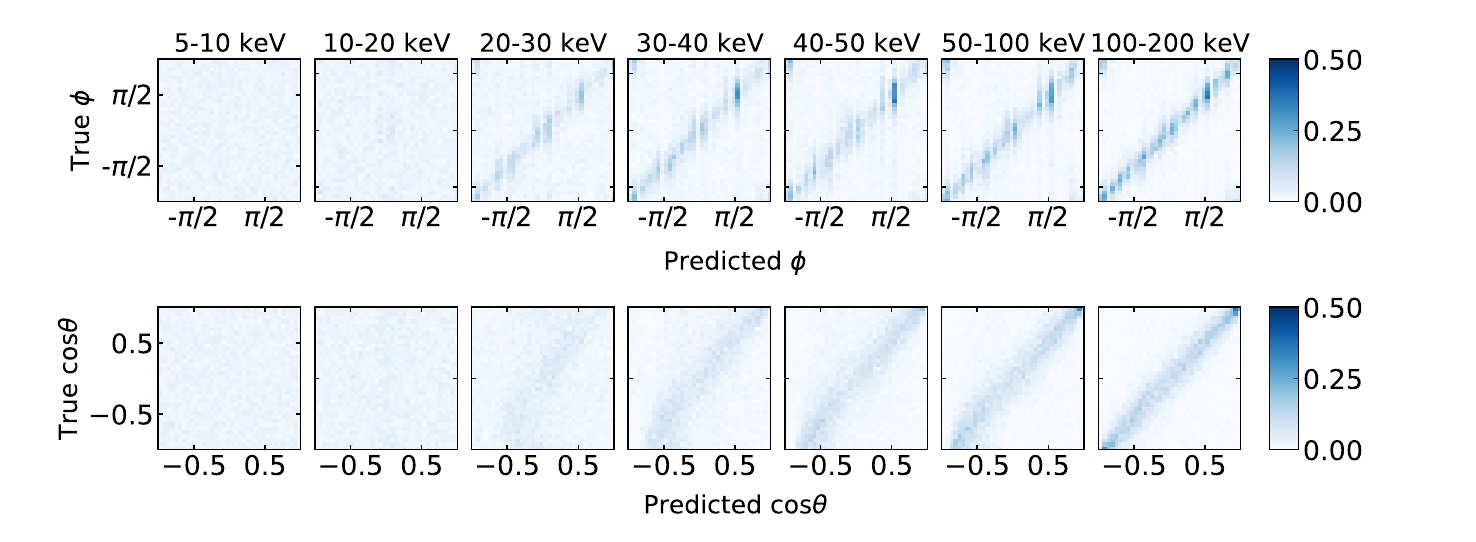}
\caption{Top panel shows the correlation matrices between the calculated angle $\phi$ and the true angle $\phi$. Bottom panel is the correlation matrices between the calculated $\cos\theta$ and the true $\cos\theta$}
\label{fig:predict_matrix}
\end{figure}

%%%%%%%%%%%%%%%%%%%%%%%%%%%%%%%%%%%%%%%%%%%%%
% Confusing matrix
%%%%%%%%%%%%%%%%%%%%%%%%%%%%%%%%%%%%%%%%%%%%%
Figure~\ref{fig:confusing_matrix} shows confusion matrices at the $\phi$ and $\cos\theta$ class predictions of several energies.In 5--10~keV of $\phi$, CNN almost misclassifies and outputs two indexes in any true $\phi$. We confirmed that such a high bias prediction appears when the highest predicted probability is very low. Therefore, the angle $\phi$ is determined by random uniformly in such a case. As the highest predicted probability exceeds a certain threshold, the predicted angle $\phi$ is calculated as follows:
\begin{equation}
\phi = {\rm atan2} \left( \sum_{i=0}^{N_{c}-1} p_{i} \sin\phi_{i},  \sum_{i=0}^{N_{c}-1} p_{i} \cos\phi_{i} \right),
\end{equation}
where $p_{i}$ is the predicted probability in $i$-th class, $\phi_{i}$ is the center angle of the $i$-th class and $N_{c}$ is the number of class. The correlation matrix between the calculated angle $\phi$ and the true angle $\phi$ are shown in Fig.~\ref{fig:predict_matrix}. The high bias prediction in 5--10~keV was reduced due to the uniform random. On the other hand, some bias prediction or the nonuniform angular response remains in 10--200~keV. In order to remove the nonuniform response, adding the nonuniform penalty term to the loss function is valuable as discussed in Ref.~\cite{KITAGUCHI}. We will deploy such an advanced method in our future work. The angle $\theta$ is calculated by the same way. The confusion matrices and correlation matrices of $\cos\theta$ is described in the bottom panels of Fig.~\ref{fig:confusing_matrix} and Fig.~\ref{fig:predict_matrix}, respectively.

%%%%%%%%%%%%%%%%%%%%%%%%%%%%%%%%%%%%%%%%%%%%%
% Evaluating Network
%%%%%%%%%%%%%%%%%%%%%%%%%%%%%%%%%%%%%%%%%%%%%
We define the two evaluated values. First one is the angular error of $\cos\theta_{\rm err} = \vec{V}_{t} \cdot \vec{V}_{p}$, where $\vec{V}_{t}$ and $\vec{V}_{p}$ are the true recoil vector and the predicted recoil vector, respectively. Second one is the position error of $R_{err}=|\vec{P}_{t}-\vec{P}_{p}|$, where $\vec{P}_{t}$ and $\vec{P}_{p}$ are the true scattering position and the predicted scattering position, respectively. We obtained these distributions by calculating for 70k validating data.
The angular and position resolutions are defined by these 50\% area value.
In the traditional method, these values are calculated in the same way.

\begin{figure}[!h]
\centering
\includegraphics[width=6in]{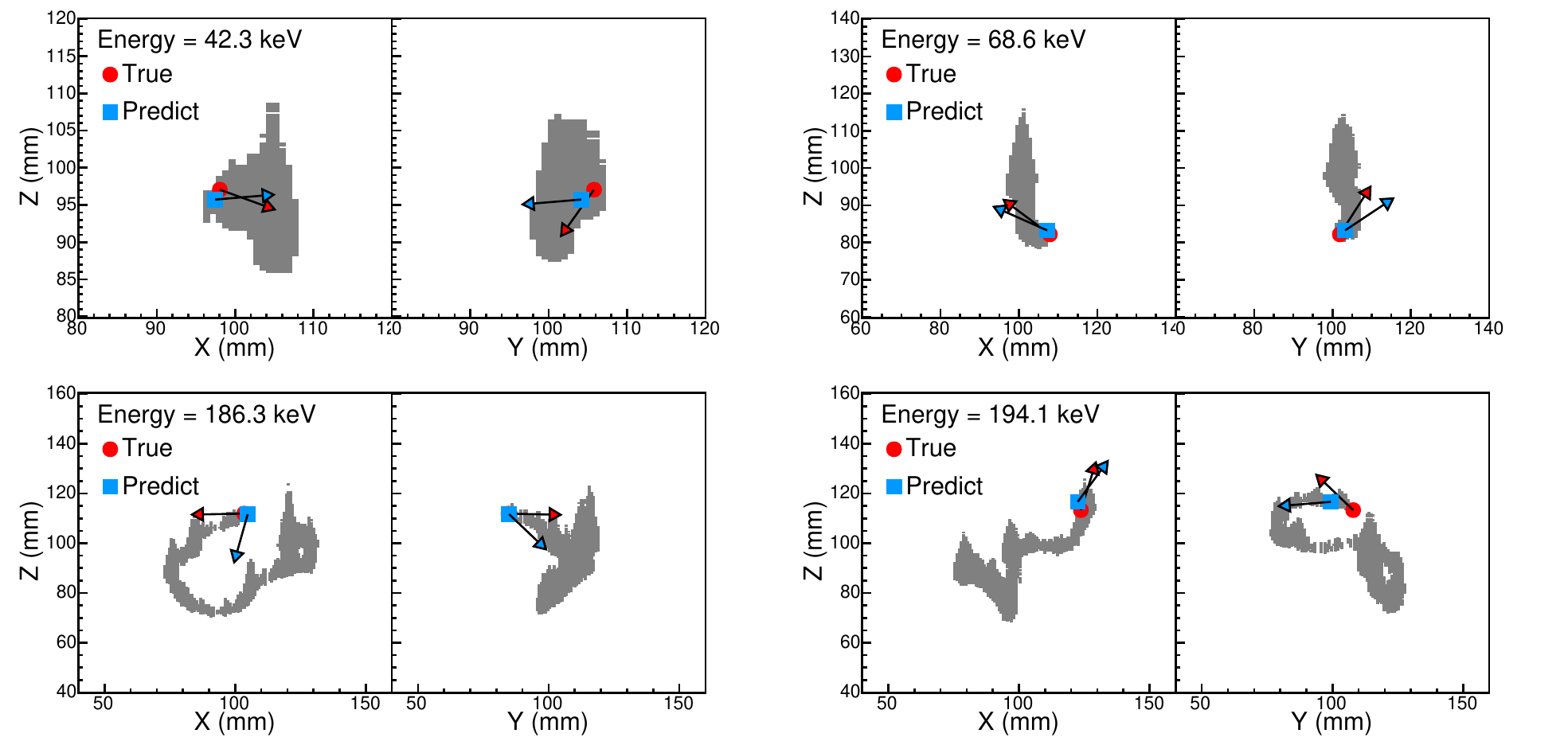}
\caption{Four track images with the predicted scattering positions and recoil directions. Red points and arrows are true scattering points and recoil directions. Blue squares and arrows are predictions using the CNN.}
\label{fig:track_sample}
\end{figure}

The several reconstructed examples on images are depicted in Fig.~\ref{fig:track_sample}. Using  the traditional method, the high-energy recoil events like the bottom figures could not reconstruct with high precision due to the multiple scattering effect. On the other hand, the CNN succeeded in resolving the scattering point and recoil direction for such events.

\begin{figure}[!h]
\centering
\includegraphics[width=6in]{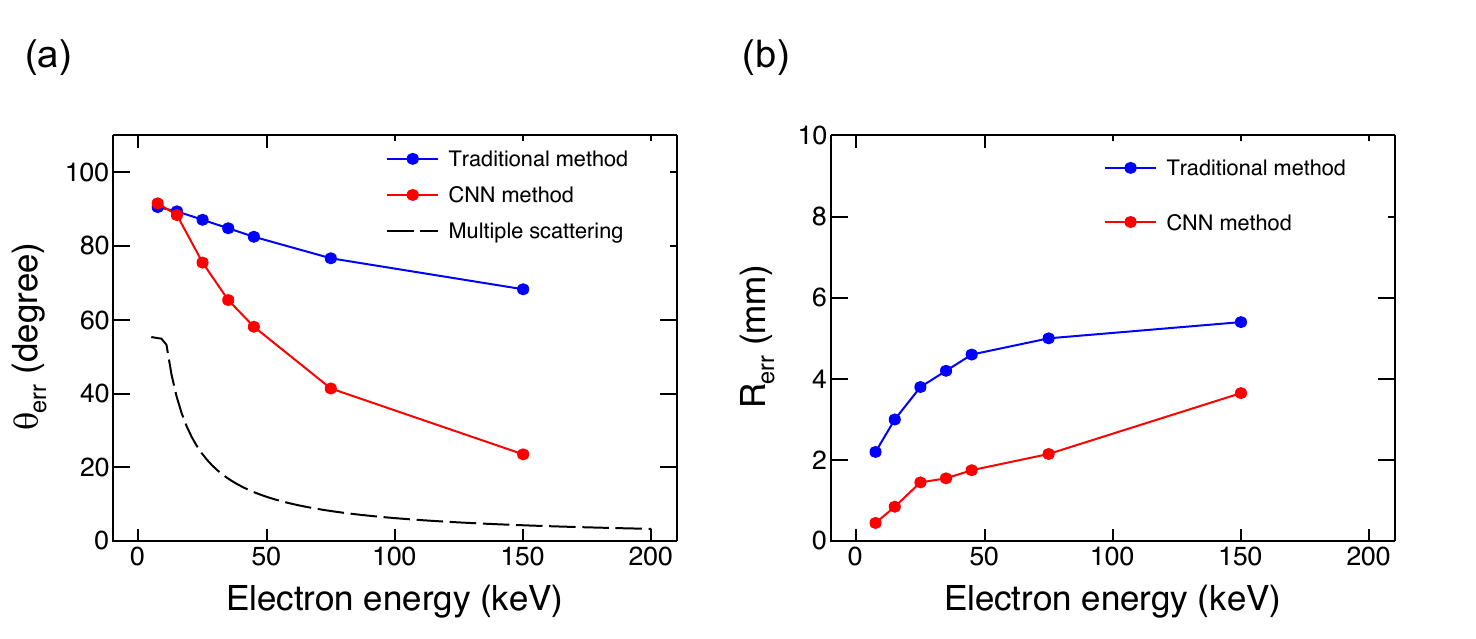}
\caption{(a) Angular resolution of the CNN method comparing to the traditional method using the simulation data. Red and blue points indicate the CNN method and the traditional method, respectively. Black dashed line is a principle limit of multiple scatterings with 0.8~mm pitch readout.
(b) Position resolution of the CNN method comparing to the traditional method. Red and blue point are the CNN method and the traditional method, respectively.}
\label{fig:Predict}
\end{figure}

Figures~\ref{fig:Predict}(a) and (b) show the energy dependence of the angular resolution and the position resolution comparing to the traditional method. 
The CNN method is better than the traditional method above 20 keV. In particular, it achieved excellent performance above 50 keV. The angular resolution of 75~keV is 41~degrees.
For predicting the scattering position, the CNN method indicates better performance in all energy. The position resolution of 75~keV is 2.1~mm.

%%%%%%%%%%%%%%%%%%%%%%%%%%%%%%%%%%%%%%%%%%%%%
% 
%%%%%%%%%%%%%%%%%%%%%%%%%%%%%%%%%%%%%%%%%%%%%
\section{Performance of ETCC using the CNN method}
We conducted the calibration measurement of ETCC for the SMILE-2+ instrument on the ground. In order to investigate several energy response, calibration data of $^{133}$Ba (mainly 356~keV), $^{22}$Na (511~keV), $^{137}$Cs (662~keV) and $^{60}$Co (1173~keV and 1332~keV) were taken. In addition, the $^{137}$Cs source was placed on several zenith positions separate to 183~cm from the center of ETCC. We reanalyzed these calibration data with the CNN method and evaluated the performance of the experimental ETCC data.

\subsection{Event selection}
In the calibration measurement, there are not only direct incident gamma-rays from the source but also backgrounds of ambient gamma-rays and cosmic-rays. Therefore, we adopted the following event selection to extract direct incident gamma-rays from the sources.
\begin{itemize}
    \item 1 Hit event : We require one track in the TPC and one hit pixel signal in the PSA, which removed most accidental coincidence events.
    \item Fiducial volume : We require that an  electron track is contained within the fiducial volume of $25.72\times25.72\times29.8$~cm$^{3}$. At the boundary of the detection area, the electric field is distorted due to the supply voltage of the PMT of the PSA around the TPC. Such events which do not retain the original track information are eliminated.
    \item Fully contain electron : The $dE/dx$ distribution or the correlation between the track-length and the energy deposit of the track in the TPC distinguishes the fully contained electrons, escaped electrons from the TPC, and cosmic-ray muons. The $dE/dx$ of the fully contained electron is empirically proportional to $E^{1.72}$~\cite{Sauli} of the recoil electron energy. We determined the fully contained electron band by fitting with Gaussian function every energy bin with a width of 10~keV, and the region within the $\pm3\sigma$ quantiles was chosen. More details can be found in Ref~\cite{Tanimori2015}.
    \item Compton kinematic test : The angle between the recoil electron direction and the scattering gamma direction is defined as Compton $\alpha$ angle and geometrically calculated by the following equation,
    \begin{equation}
        \cos\alpha_{\rm geo} = \vec{g}\cdot \vec{e}.
    \end{equation}
    In addition, this angle is derived by the Compton scattering kinematics:
    \begin{equation}
        \cos\alpha_{\rm kin} = \left(1-\frac{m_{e}c^{2}}{E_{\gamma}}\right)\sqrt{\frac{K_{e}}{K_{e}+2m_{e}c^{2}}}.
    \end{equation}
    Therefore, $\Delta\cos\alpha$~($=\cos\alpha_{\rm geo} - \cos\alpha_{\rm kin}$) nearly zero ensures that the reconstructed events are true Compton-scattering events. We limited as  $|\Delta\cos\alpha|<0.5$ in order to extract Compton scattering events.
    \item Energy selection : Events, which reconstructed energy is within two times of FWHM for the source energy, are used to extract the direct components.
\end{itemize}

After the event selection, the effective area of experiment and simulation data for 662~keV were $2.3\times10^{-1}$~cm$^{2}$ and $2.1\times10^{-1}$~cm$^{2}$, respectively.

%%%%%%%%%%%%%%%%%%%%%%%%%%%%%%%%%%%%%%%%%%%%%%%%%%%%%%%%%%
\subsection{Imaging performance}\label{sec:imaging}
\begin{figure}[!h]
\centering
\includegraphics[width=6in]{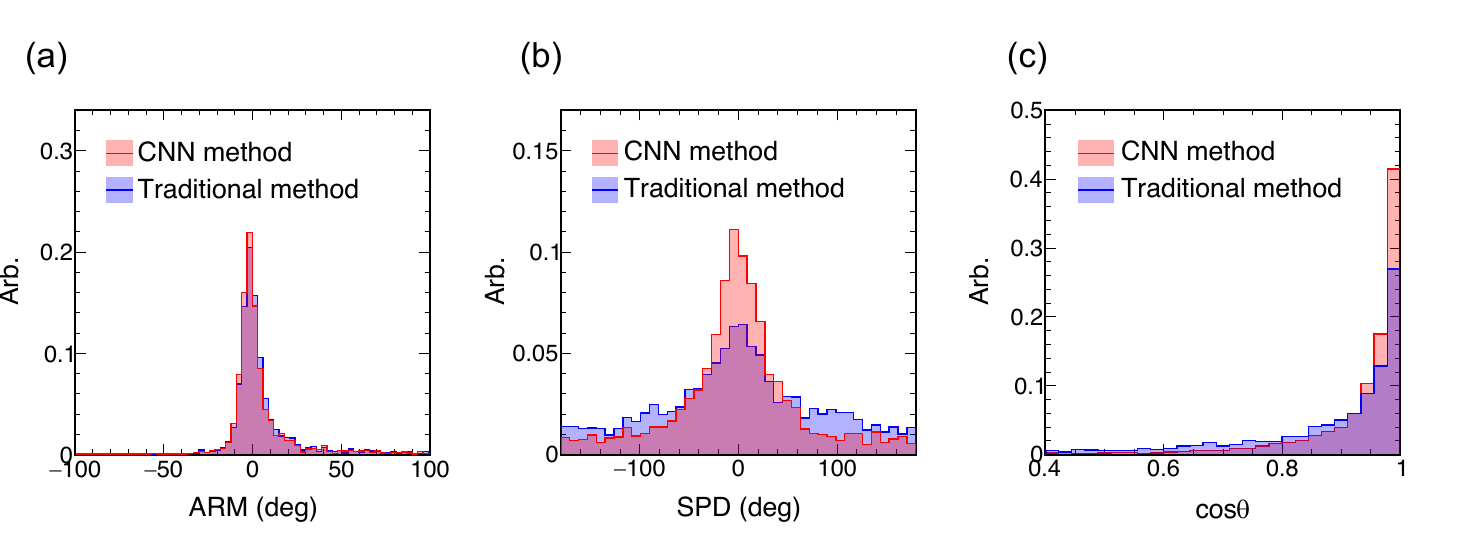}
\caption{
(a)~ARM, (b)~SPD and (c)~$\cos\theta$ distributions of the experiment data. Red and Blue histograms, which are normalized by the total areas, indicate the CNN and the traditional method, respectively.}
\label{fig:ARMSPDPSF}
\end{figure}

The incident gamma-ray direction is reconstructed by Eq.~(\ref{eq:reconstruct}).
To evaluate the reconstruction accuracy, we defined the error angle  $\cos\theta$ between the reconstructed vector and the source vector. The PSF is the angular resolution of $\theta$. Thus, we calculated the half area value of the $\cos\theta$ distribution as the PSF.
Figure~\ref{fig:ARMSPDPSF} shows the ARM, SPD, and $\cos\theta$ distribution using the CNN method with the traditional method as the $^{137}$Cs calibration source was set at the zenith=0.
%The ARM of CNN method is 10 degrees and not different from the traditional method, since the PSA position uncertainty is dominated the ARM.
The ARM using the CNN method is 10~degrees and is not different from the traditional method. We confirmed that the supply voltage of PSAs around TPC distorted the electric field, and it dominated the uncertainty of the scattering position.
On the other hand, the SPD was highly improved thanks to reducing the angular resolution of the recoil direction. We obtained 59 degrees (FWHM) of SPD by fitting Gaussian function. This benefit gave the sharp $\cos\theta$ distribution and improved the PSF to 15 degrees. 

\begin{figure}[!h]
\centering
\includegraphics[width=6in]{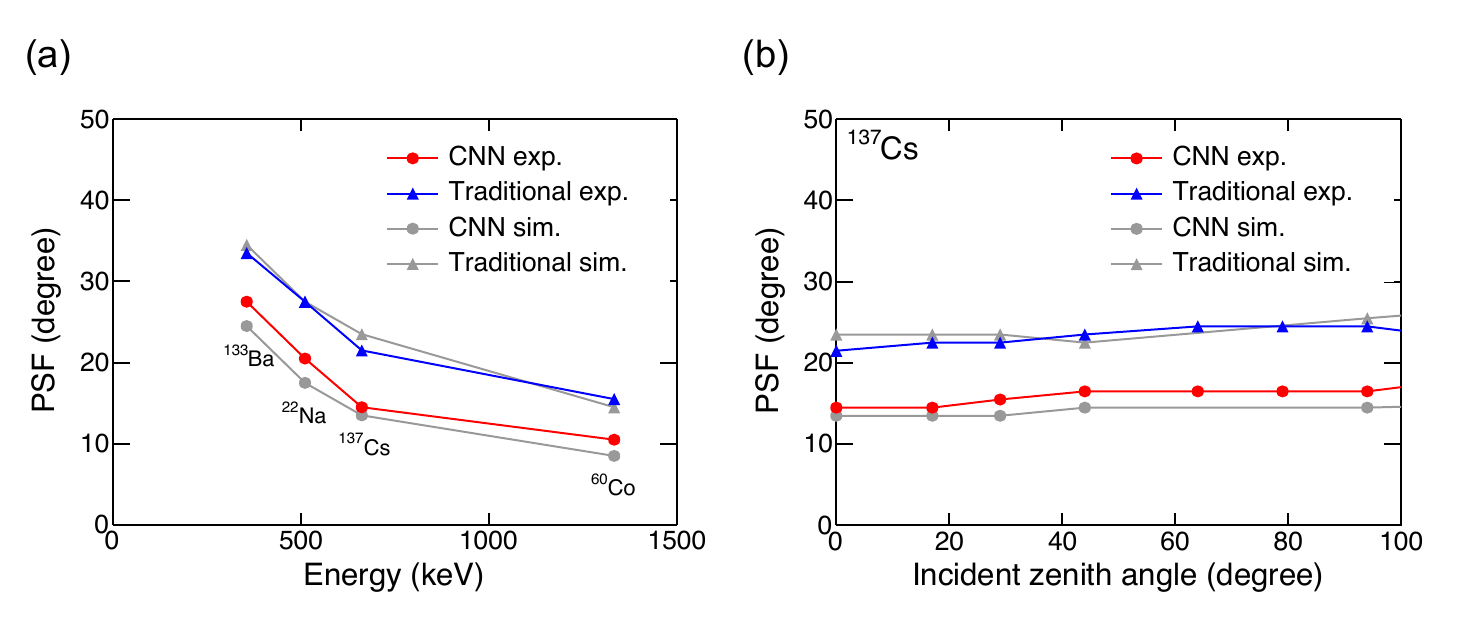}
\caption{(a)~Gamma-ray energy dependence and (b) incident zenith-angle dependence of PSF. Red and blue lines indicate the experiment data using the CNN method and the traditional method, respectively. Gray dots and triangles are the simulation data using the CNN method and the traditional method, respectively.}
\label{fig:psf_compare}
\end{figure}

Figure~\ref{fig:psf_compare}(a) and (b) show the dependencies of PSF on gamma-ray energy  and the incident zenith-angle, respectively. For every gamma-ray energy and incident zenith-angle, the CNN method marked better performance than the traditional method.
% Again, the main reason is that the SPD or the angular resolution improved.

\begin{figure}[!h]
\centering
\includegraphics[width=6in]{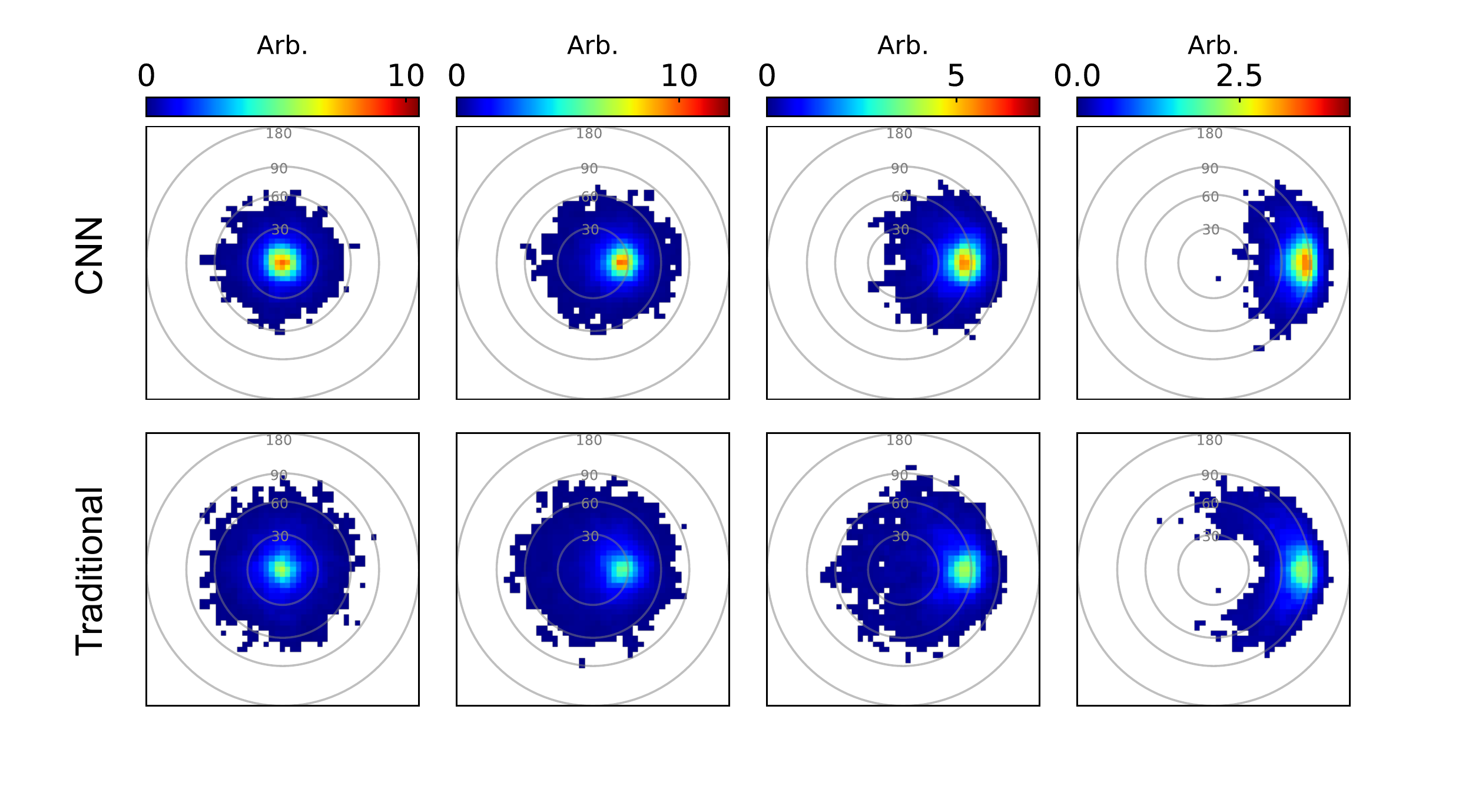}
\caption{Reconstructed source images of $^{137}$Cs set at zenith angles of 0, 30, 60, and 90 degrees for the experiment data. The top and bottom columns are the CNN and traditional methods, respectively.}
\label{fig:Recon_image}
\end{figure}

Reconstructed source images of $^{137}$Cs for several zenith angles are shown in Fig.~\ref{fig:Recon_image} using equal-solid-angle projection, such as Lambert projection.
The bright points representing sources were observed, and the more focused images were confirmed in the CNN method.

%%%%%%%%%%%%%%%%%%%%%%%%%%%%%%%%%%%%%%%%%%%%%
% Discussion 
%%%%%%%%%%%%%%%%%%%%%%%%%%%%%%%%%%%%%%%%%%%%%
\section{Discussion}
Since the scattering position resolution is smaller than the PSA position uncertainty, further improvement in scattering points cannot be expected to improve PSF. On the other hand, SPD remains improvement potential. Although the recoil direction was dramatically improved using CNN, the physical limit of the multiple scattering has not been reached yet. As the angular resolutions of the azimuth and the zenith angles were calculated separately for 50~keV electrons at the 4~mm distance from the scattering points, 29~degrees and 35~degrees were obtained, respectively. This indicates that the zenith angular resolution mainly determines a 3-D angular resolution.  The Z-direction position resolution has a width corresponding to the TOT. From a rough calculation, the drift speed of 38~mm/$\mu$s and TOT width of 80~ns corresponding to 0.5~pC input charge give 3.0~mm position resolution. It is 3.8 times worse than the 800~$\mu$m read-out pitch of X and Y directions. This is the cause of smearing the track information for low energy electrons, as can be seen from the image of 42.3~keV in the upper left of Fig.~\ref{fig:track_sample}.
Improvement can be expected by using a short shaping time amplifier. % and fine pixel intervals.
Also, waveforms taken by the flush ADC have the Z-direction information~\cite{DRIFT}. Therefore, utilizing the waveform images in CNNs would be a good approach.
Regarding the azimuth direction, it is adequate to make the read-out pitch more acceptable and increase the track information using such as the three projections read-out system~\cite{HexaSauli,DUNE}. We confirmed that the three-projections read-out system with 480~$\mu$m pitch achieved 50~degrees of the 3-D angular resolution from the simulation.

In this research, a network based on VGG was used as a demonstration. However, residual network~\cite{ResNet} (ResNet) is mainly used to realize a deeper layer in recent research. In the future, we will incorporate this technology to improve the neural network aspect.

%%%%%%%%%%%%%%%%%%%%%%%%%%%%%%%%%%%%%%%%%%%%%
% SUMMARY
%%%%%%%%%%%%%%%%%%%%%%%%%%%%%%%%%%%%%%%%%%%%%
\section{Conclusion}
We designed two CNNs based on the VGG and U-Net models to predict the scattering position and the electron-recoil direction from track images taken by ETCC. The angular resolution and scattering position resolution utilizing CNNs are obtained as 41 degrees and 2.1 mm for 75 keV in simulation data. As we adopted the CNN analysis to experimental calibration data, the resulting PSF was improved by 32\% compared with the traditional analysis and achieved 15~degrees. 
The CNN analysis surpassed the traditional analysis, and such improved ETCC has the potential to reveal the MeV gamma-ray sky.

%%%%%%%%%%%%%%%%%%%%%%%%%%%%%%%%%%%%%%%%%%%%%%%%%%%%%%%%%%%%%%%%%%
%%%%%%%%%% ACKNOWLEDGMENT %%%%%%%%%%%%%%%%%%%%%%%%%%%%%%%%%%%%%%%%
%%%%%%%%%%%%%%%%%%%%%%%%%%%%%%%%%%%%%%%%%%%%%%%%%%%%%%%%%%%%%%%%%%
\section*{Acknowledgment}
This work was partly supported by JSPS (Japan Society for the promotion of Science) KAKENHI (Grant-in-Aids for Scientific Research) (grant nos. 20K20428).

%%%%%%%%%%%%%%%%%%%%%%%%%%%%%%%%%%%%%%%%%%%%%%%%%%%%%%%%%%%%%%%%%%
%%%%%%%%%% REF %%%%%%%%%%%%%%%%%%%%%%%%%%%%%%%%%%%%%%%%%%%%%%%%%%%

\bibliographystyle{ptephy}
\bibliography{main}
%
% once the .bbl file has been generated then place the text in your article.

\end{document}